\documentclass[12pt,preprint]{aastex}

\begin{document}

\title{Formation of stars and clusters over cosmological time}

\author{Bruce G. Elmegreen}
\affil{IBM Research Division, T.J. Watson Research Center, 1101 Kitchawan
Road, Yorktown Heights, NY 10598, bge@watson.ibm.com}

\begin{abstract}
The concept that stars form in the modern era began some 60 years ago with
the key observation of expanding OB associations. Now we see that these
associations are an intermediate scale in a cascade of hierarchical
structures that begins on the ambient Jeans length close to a kiloparsec in
size and continues down to the interiors of clusters, perhaps even to
binary and multiple stellar systems. The origin of this structure lies with
the dynamical nature of cloud and star formation, driven by supersonic
turbulence and interstellar gravity. Dynamical star formation is also
relatively fast compared to the timescale for cosmic accretion, and in this
limit of rapid gas consumption, the star formation rate essentially keeps
up with the accretion rate, whatever that rate is, leading to a sequence of
near-equilibrium states during galaxy formation and evolution. In these
states, most of the large-scale and statistical properties of galaxies and
of star formation inside them depend on the rate of cosmic accretion, which
is a function of galaxy mass and epoch. This simplification of star
formation on a cosmic scale allows successful simulations and modeling even
when the details of star formation on the scale of molecular clouds is
unclear.  Dynamical star formation also helps to explain the formation of
bound clusters, which require a local efficiency that exceeds the average
by more than an order of magnitude. Efficiency increases with density in a
hierarchically structured gas. Cluster formation should vary with
environment as the relative degree of cloud self-binding varies, and this
depends mostly on the ratio of the interstellar velocity dispersion to the
galaxy rotation speed.  As this ratio increases, galaxies become more
clumpy, thicker, and have more tightly bound star-forming regions.  The
formation of old globular clusters is understood in this context, with the
metal-rich and metal-poor globulars forming in high-mass and low-mass
galaxies, respectively, because of the galactic mass-metallicity relation.
This relation is a result of the accretion equilibrium.  Metal-rich
globulars remain in the disks and bulge regions where they formed, while
metal-poor globulars get captured as parts of satellite galaxies and end up
in today's spiral galaxy halos.  Blue globulars in the disk could have
formed very early when the whole Milky Way had a low mass.
\end{abstract}

\section{Celebrating the Beginnings: 60 and 64 years ago}

A good place to begin this story is with the publications by Viktor
Ambartsumian in 1949 and 1950, and then again in the Western literature in
1954, of his discovery that luminous stars are in loose associations that
expand over time. This was the beginning of the modern era of
star-formation research. The only explanation for the systematic motion was
that the associations are young, with expansion ages of several million
years, and therefore that the stars themselves are young. Before this,
astronomers could not really tell if star formation was still going on in
the present-day universe. The age of the solar system, galactic stars, and
the Hubble time were all thought to be about 3 Gyr (see 1935 {\it Science}
Volume 82). \cite{blaauw52} quickly confirmed Ambartsumian's discovery.
Prior speculation about the mechanisms and locations of star formation
\citep{bok36, edgeworth46, whipple46, spitzer49}, whether it took place in
the early universe or more recently (as suggested to Spitzer by the high
luminosities of supergiant stars), suddenly fell into place.

A flurry of activity followed. In 1952 Herbig speculated that the variable
stars he had been studying in nebulae could actually be young and not just
interacting with nearby gas. \cite{opik53} proposed that stellar
associations expanded because of supernova-induced star formation, and
\cite{zwicky53} proposed that the expansion came from cluster unbinding as
the gas left. \cite{morgan53} discovered spiral arms in the Milky Way from
the positions of OB associations.  \cite{hoyle53} proposed that collapsing
clouds would fragment, making clusters. Soon after, \cite{salpeter55}
determined the distribution function of stellar masses at intermediate mass
in the solar neighborhood. Also at about this time, \cite{ewen51} and
\cite{muller51} discovered the 21 cm emission line of hydrogen, after which
\cite{lilley55} determined the gas-to-dust ratio, and \cite{bok55} found
that dark clouds had too little HI for their dust, suggesting that the gas
is molecular.

It is interesting to note that a possible impediment to early recognition
that the interstellar medium is gravitationally unstable to collapse into
clouds and stars was the wrong gas-to-dust ratio; the interstellar gas mass
was thought to be too low by a factor of $\sim10$ \citep[see][page
243]{spitzer41}. This was before interstellar HI was detected, and,
although heavy elements like sodium and calcium were observed in
interstellar absorption lines, it was also before the extent of their
depletion onto grains was known. \cite{oort32} showed from vertical gravity
that the non-stellar mass amounted to an average density of about 1 atom
cm$^{-3}$, but the form of this matter was not known (he included
meteorites in his possible list of mass contributors). Thus
\cite{spitzer41} and \cite{whipple46} tried to make stars by the
convergence of radiation pressure on mutually shielding dust particles,
which had a 1/2 Gyr timescale in the ambient medium. The idea that a medium
with an average density of $\sim1$ atom cm$^{-3}$ could be gravitationally
unstable on an enormous scale, forming kiloparsec-size cloud complexes with
a cascade to smaller scales forced by supersonic turbulence, had to wait
for a more general understanding of spiral structure and large-scale disk
dynamics, and for extensive observations of correlated gas structures and
motions. \citep[For a review of the contributions by Lyman Spitzer Jr. to
the early developments of star formation theory, see][]{elmegreen09}.

A personal milestone was the IAU Symposium 76 on ``Star Formation,'' held
in Geneva Switzerland in 1976. This marked the year of my first publication
on this topic. It was also a good place for a honeymoon. In the context of
the present talk, the conference summary by Donald Lynden-Bell (1977) was
noteworthy. There he said: ``There are three great areas of astronomy to
which a theory of star formation is vital,'' and the first area listed was
``The formation and evolution of galaxies.''  For this he described a
complicated hypothetical star formation rate ${\cal R}$ that is a function
of many environmental variables but then said ``However, if we knew the
true functional form of ${\cal R}$ and offered it to a galaxy builder he
would probably tell us `Oh, go jump in the lake, that's far too
complicated'. Thus galaxy builders need oversimplified average laws like
Schmidt's suggestion ${\cal R}=C\rho^2$.'' After much research since then
by a large community on star formation triggering in pillars, shells,
spiral arms, and supersonically compressed turbulent clouds, we have come
back to Lynden-Bell's suggestion.

\section{The Oversimplified Average Law}
\label{oversim}

``Oversimplified average laws'' like the Schmidt law, or, in today's terms,
the Kennicutt-Schmidt law, have indeed become the preferred form for galaxy
builders.  In fact, the star formation law can be even simpler, i.e.,
linearly proportional to total galactic gas mass, and even then most of the
statistical properties of galaxies can be derived without regard to
detailed processes on stellar scales. These properties include the
distribution functions of stellar, gaseous, and metal masses over cosmic
time, along with the star formation rates. The reason for this
simplification is that the star formation timescale in terms of the time to
use up the available gas is generally shorter than the timescale for cosmic
evolution. Then star formation keeps up with the accretion rate and all of
these large-scale distribution functions depend primarily on the properties
of cosmic accretion \citep{larson72,edmunds90,dave12,lilly13,dekel13,
peng14}.

Following \cite{dekel13}, the baryonic accretion rate through the galaxy
virial radius at redshift $z$ is
\begin{equation}
{{dM_{\rm acc}}\over{dt}}=30f_bM_{\rm halo,0,12}e^{-0.79z}(1+z)^{2.5}\;
M_\odot \;{\rm yr}^{-1},
\end{equation}
where $M_{\rm halo,0,12}$ is the halo mass today in units of
$10^{12}\;M_\odot$ (equal to about 2 for the Milky Way), and the baryonic
fraction is $f_b=0.16$. The accretion rate to the disk could be about the
same as the rate through the virial radius when the halo mass is equal to
or less than $10^{12}\;M_\odot$, because then accretion mostly penetrates
the halo in a cold flow \citep{keres05,dekel06}. For a review of accretion
and star formation, see \cite{sanchez14}.

The star formation rate in these simple models is
\begin{equation}
{\rm SFR} =  M_{\rm gas}/\tau
\label{SFR}
\end{equation}
for gas consumption time $\tau$ that may depend on stellar or halo mass and
redshift.  Gas is also lost to the halo in a wind at the rate
\begin{equation}
{{dM_{\rm gas}}\over{dt}}=-w \times {\rm SFR}
\end{equation}
where $w$ is the mass loading factor for winds \citep[$w\sim1$
in][]{peng14}; $w$ may also depend on galaxy mass and redshift
\citep{zahid14}. If stars return a fraction $R$
\citep[$\sim0.4$;][]{peng14} of their mass to the interstellar medium, then
the interstellar mass varies with time as
\begin{equation}
{{dM_{\rm gas}}\over{dt}}={{dM_{\rm acc}}\over{dt}}-(1-R+w)M_{\rm gas}/\tau .
\end{equation}
If $\tau$ is small compared to timescales for changes in the accretion
rate, then there is a steady state where $dM_{\rm gas}/dt\sim0$ and
\begin{equation}
{\rm SFR} \sim \left(1\over{1-R+w}\right){{dM_{\rm acc}}\over{dt}}.
\end{equation}

\cite{dekel13} show that these equations compare well to numerical
simulations. They also derive the consumption time $\tau$ to be about
$0.17t_{\rm H}$ for instantaneous Hubble time $t_{\rm H}$. This time
relation comes from the ratio of the virial radius to the virial speed,
which depends on $t_{\rm H}$, combined with the ratio of the disk radius to
the virial radius, which comes from the halo spin parameter, plus the
similarity between the free fall time at the average gas density and the
disk dynamical time, and the efficiency of star formation per unit free
fall time.  The point is that $\tau$ is smaller at earlier times and always
small enough to keep the galaxy in a quasi-equilibrium state, even when the
galaxy is young.

Equation (\ref{SFR}) comes from a more fundamental relation
\begin{equation}
{\rm SFR/Volume} = \epsilon_{\rm ff}\rho/\tau_{\rm ff}
\label{SFR2}
\end{equation}
where $t_{\rm ff}$ is the free fall time and $\epsilon_{\rm ff}$ is the
fraction of the gas mass that turns into stars in a free fall time.
Generally $\epsilon_{\rm ff}$ is small, several per cent, for densities
close to the average interstellar density \citep{krumholz07}.
\cite{krumholz12} show that all regions of star formation, ranging from
individual molecular clouds to normal galaxies to starburst galaxies to
high-redshift galaxies, satisfy this relation if the free fall time is
taken to be the minimum of two values, the value for giant molecular
clouds, which have much higher densities than the average ISM in normal
disks, and the value for the ambient medium, with the additional conditions
that the gas pressure is determined by the weight of the mass column
density and the Toomre instability parameter is constant and of order
unity.

\section{Why is $\epsilon\sim0.01$?}

The low value of $\epsilon_{\rm ff}\sim0.01$ suggests a broad spectrum of
possibilities, ranging from molecular clouds and other interstellar gas
that is prevented from collapsing for several tens of free fall times, to
the presence of only a small fraction of the gas in a form that can
collapse into a star, with a rapid collapse rate for this small fraction.
The observation of bound star clusters that form locally in only several
free fall times \citep{getman14} suggests that $\epsilon_{\rm ff}$ is
actually high in dense regions, because a high total efficiency is
necessary to keep the stars bound to the cluster after the gas leaves
\citep{parmentier09}. In this case $\epsilon_{\rm ff}$ should be an
increasing function of density. However, such an increase is inconsistent
with the expected speed-up of collapse at high density: $\epsilon_{\rm
ff}M/\tau_{\rm ff}$ in a region of total mass $M$ should give the same star
formation rate whether it is measured at high density or low density.
Because $\tau_{\rm ff}\propto\rho^{1/2}$, $\epsilon_{\rm ff}$ should scale
as $\rho^{-1/2}$ to keep $\epsilon_{\rm ff}M/\tau_{\rm ff}$ constant if $M$
is constant.

A solution to this problem \citep{elmegreen02} is that only a low fraction
of the total mass $M$ can be at a high density, collapsing rapidly on its
local dynamical time, while a much larger fraction of $M$ is at a low
density, collapsing slowly or unable to collapse at all on its own
dynamical time. The collapsing fraction has to decrease with increasing
average density. Thus $\epsilon_{\rm ff}$ should be viewed as including a
term equal to the fraction of the gas at a density greater than $\rho$
\citep{elmegreen08}. Then the star formation rate is independent of the
scale of measurement because
\begin{equation}
\epsilon_{\rm ff}(\rho)\rho^{1/2}f_{\rm M}(\rho) =
\epsilon_{\rm ff}(\rho_0)\rho_0^{1/2}f_{\rm M}(\rho_0)
\end{equation}
for either some normalization at an average density $\rho_0$, or, what is
more physically motivated, a normalization at a some high density $\rho_0$
where the local efficiency has its maximum possible value close to unity.
Here $f_{\rm M}(\rho)$ is the mass fraction at density larger than $\rho$.
Depending on the details of collapse and cloud structure, we could
alternatively write the above equation as
\begin{equation}
\epsilon_{\rm ff}(\rho)\int_\rho \rho^{1/2}f^\prime_{\rm M}(\rho)d\rho =
\epsilon_{\rm ff}(\rho_0)\rho_0^{1/2}f_{\rm M}(\rho_0)
\end{equation}
where $f^\prime$ is the differential form of $f$, namely the fraction of
the gas between density $\rho$ and $\rho+d\rho$. As a result of these
considerations, the local efficiency increases with $\rho$ as
\begin{equation}
\epsilon_{\rm ff}(\rho)=\epsilon_{\rm ff}(\rho_0)
\left({{\rho_0}\over{\rho}}\right)^{1/2} {{f_{\rm M}(\rho_0)}
\over{f_{\rm M}(\rho)}}.
\end{equation}

The mass fraction $f_M(\rho)$ is most likely determined by turbulent
fragmentation \citep{elmegreen02,elmegreen08}, which makes this function a
log-normal in weakly self-gravitating regions or a log-normal with a
power-law tail in strongly self-gravitating regions
\citep{klessen00,vs08,kritsuk11,elmegreen11}. Turbulent fragmentation
partitions a cloud hierarchically so that dense regions are inside and to
the side of lower-density regions \citep{burkhart13}. With hierarchical
structure, the fraction of the mass in the highest density regions
increases as the average density increases (Figure 1). If we consider that
the formation of a bound cluster requires a certain minimum local
efficiency, like $\sim10-20$\%, depending on the timescale for gas clearing
(it has to be 50\% in the immediate neighborhood of the stars for rapid
clearing), then this minimum local efficiency corresponds to a minimum
average local density in comparison to the fiducial density $\rho_0$
\citep{elmegreen08, elmegreen11}. Thus star formation at low average
density, as in star complexes \citep{efremov95} or OB associations, has a
low efficiency, several percent, and leaves the stars expanding, as found
by Ambartsumian 64 years ago. Star formation at high average density,
inside molecular cloud cores, for example, has a high efficiency, 10\% or
more, often leaving a bound cluster. At even higher density, where
individual or multiple stellar systems form, the efficiency can approach
50\% or more. There is no characteristic length or mass scale for star
formation (aside from the lower mass limit to the power law in the initial
stellar mass function). Stellar groupings form with a wide range of
densities, following the log-normal distribution \citep{bressert10}.

This simplified view of bound cluster formation glosses over many details,
of course, but it has predictive value and it coincides qualitatively with
the current picture of star formation in a turbulent medium.  One
prediction of it is that the density at which star formation becomes highly
efficient, perhaps viewed as a critical density for star formation or a
density at which self-gravity becomes stronger than other forces, should
increase in regions with higher Mach number and/or higher average density.
The critical density should be viewed as relative to the average rather
than as absolute. This prediction comes from the efficiency function
$\epsilon(\rho)$, which, for a log-normal density pdf and mass function
$f_{\rm M}(\rho)$, reaches a certain high value when the density exceeds a
certain multiple of the average. There is no absolute scale in this
argument. Moreover, the ratio of the density at high efficiency to the
average density increases with both Mach number \citep{elmegreen08} and
central concentration of the cloud \citep{elmegreen11}. Higher pressures
should therefore correspond to higher threshold densities, a higher
fraction of star formation in bound clusters, and denser clusters.
Applications of this scaling to the central molecular zone of galaxies were
made by \cite{kruijssen14}, who explained the modest star formation rate
there (considering the $\times100$ density compared to the disk) as a
result of a higher threshold density.

Another application is to clumpy galaxies, such as high redshift clumpy
galaxies and local dwarf irregulars.  As discussed more in Sections
\ref{clumpy} and \ref{ssc} below, clumpy galaxies have high ratios of
velocity dispersion to rotation speed. In this case the relative binding
energy of the star-forming regions is high too.  Such high binding energy
corresponds to more strongly concentrated clouds and a higher mass fraction
forming stars at high efficiency, according to the discussion above. Thus
we predict a higher fraction of star formation in bound clusters, and
possibly denser clusters, in clumpy galaxies than in smooth-disk galaxies.

Also important in the determination of $\epsilon$ is the geometry of the
star-forming gas at high density, which tends to be filamentary
\citep[e.g.][]{hacar13,palmeirim13}.  Filamentary gas drains down to cores
on a timescale equal to about the dynamical time, $(G\rho)^{-1/2}$, for
local filament density $\rho$, multiplied the ratio of the length to width
of the filament. This extra multiplicative factor is the result of gravity
being an isotropic central force with a dynamical time proportional to the
inverse square root of the average density inside an enclosing sphere,
rather than the inverse square root of the local density inside sub-regions
of that sphere. Thus there is a slowness to filamentary star formation
compared to spherical star formation at the same local (peak) density. This
means that the accreting core can sustain prestellar growth for many local
dynamical times as the attached filament or filaments drain into it. Still
the process is dynamical, that is, without equilibrium from feedback or
delays from magnetic diffusion. It is just that the timing has a
contribution from the geometry in addition to the local peak density.

\section{Starbursts are Thick}

\cite{krumholz12} model equation (\ref{SFR2}) in a way that allows them to
convert from a volume density $\rho$ to a surface density $\Sigma$, for
comparison with observations. To do this they make two reasonable
assumptions, that the value of Toomre $Q=\kappa\sigma/\left(\pi
G\Sigma\right)$ is about constant, as determined by gravitational
instability feedback regulating the velocity dispersion $\sigma$ ($\kappa$
is the epicyclic frequency), and the gas pressure is determined by
hydrostatic equilibrium, $\rho\sigma^2\sim0.5\pi G\Sigma^2$. The first
relation says that $\sigma\propto\Sigma$ for a given $\kappa$ and then the
second says that $\rho=$ constant at that $\kappa$.  This means that the
midplane density is about the same in all galaxy disks of the same size and
rotation rate, regardless of whether the galaxy has low or high surface
brightness from star formation.

The projected star formation rate in this interpretation is $\epsilon_{\rm
ff}\Sigma/t_{\rm ff}$, and so it varies only as $\Sigma$ varies for a given
midplane density (i.e., fixed $t_{\rm ff}$), and $\Sigma\propto\sigma$, the
velocity dispersion. Thus higher star formation rate densities correspond
to higher velocity dispersions. Considering also that the galaxy thickness
varies as $H\sim\sigma^2/\left(\pi G\Sigma\right)$ for a one-component
disk, it follows that $H\propto\sigma\propto\Sigma$. Thus galaxies with a
high surface brightness of star formation are thicker on the line of sight,
and have a higher gas velocity dispersion, than galaxies with a low surface
brightness from star formation, all else being equal. Starburst galaxies
have more layers of fixed density on the line-of-sight than low-surface
brightness galaxies of the same size and rotation rate. A starburst merger
of a given size, like the Antennae galaxies, is thicker than an isolated
galaxy (and has a larger velocity dispersion).  Large HI velocity
dispersions in interacting galaxies were measured long ago
\citep{elmegreen93} and have been modeled by computer simulations
\citep{powell13}.

\section{LEGUS}

A recent HST survey of nearby galaxies in the ultraviolet, LEGUS (PI:
Calzetti, 2014, in preparation) has been analyzed to determine the
characteristics of hierarchical structure in star-forming regions
\citep{elmegreen14}. There were 12 galaxies in the analysis covering a
range of late Hubble types and star formation rates, including NGC 1705 and
NGC 5253, which contain super star clusters. The analysis consisted of
blurring the images in successively increasing sizes and counting the
number of regions in each blurred image. A log-log plot of this number
versus the blur scale typically gives a power law, and the slope of this
power law is the projected fractal dimension.

The study indicated that the fractal dimension of a whole galaxy is higher,
more like 2 than 1, in the most actively star-forming galaxies (including
NGC 1705, NGC 5253, and UGC 695) and that individual regions in all of the
galaxies also have high dimensions ($\sim2$). Inactive galaxies have low
overall dimensions ($\sim1-1.5$) although their individual regions have
high dimensions. This means that the basic building block for star
formation is a bright patch with a high dimension, and that the most active
galaxies are composed of only one or two of these patches, which
essentially cover a high fraction of the disk. The less active galaxies,
including some with spiral arms, typically have their star-forming regions
spread out, and this gives them a lower overall fractal dimension. A high
fractal dimension, $\sim2$, corresponds to a region that is filled with
emission in projection; there are relatively few empty lines of sight.

\section{Clumpy Galaxies and Super Star Clusters}
\label{clumpy}

The number of giant star-forming regions, or star complexes, in a galaxy is
an important discriminant of star formation activity.  Each region
typically has a size and mass comparable to the Jean length and mass at the
average turbulent speed and average density of the interstellar medium. The
Jeans length is essentially $\sigma/\left(G\rho\right)^{1/2}$. The galaxy
size, on the other hand, scales with the rotation speed $V_{\rm circ}$
instead of the velocity dispersion, and may be written approximately as
$V_{\rm circ}/\left(G\rho\right)^{1/2}$. The number of Jeans lengths in a
galaxy of a given average density is therefore $(V_{\rm circ}/\sigma)^2$.
This number is small when the gaseous velocity dispersion is a high
fraction of the rotation speed.  This is the case for local dwarf
Irregulars and BCDs, which have about the same $\sigma$ as local spirals
but lower $V_{\rm circ}$, and also for high redshift clumpy galaxies, which
have about the same $V_{\rm circ}$ as local spirals but much higher
$\sigma$ \citep{for09}.

The disk thickness and turbulent Jeans length are nearly the same,
$\sigma^2/\left(\pi G\Sigma\right)$, so galaxies with high $\sigma/V_{\rm
circ}$ have relatively thick disks (relative to their diameters). This is
the case for local dwarf irregulars and clumpy high-redshift disks. In the
case of the clumpy high-z disks, the relative thickness is presumably what
produced the old thick disks in today's galaxies \citep{bournaud09}.
Considering the results of the previous paragraph, the relative disk
thickness scales inversely with the number of giant star complexes.

\section{Super Star Clusters}
\label{ssc}

Now we come to the formation of super star clusters. How can some small
galaxies like NGC 1705 produce a $10^6\;M_\odot$ bound cluster when large
galaxies like the Milky Way only produce clusters up to
$\sim10^5\;M_\odot$? \citep[Some local spiral galaxies have more massive
clusters,][ but the Milky Way apparently does not.]{larsen99}. The answer
again seems to come down to the number of star complexes. The ratio of the
gravitational binding energy in a Jeans mass cloud complex to the
background energy density from the galaxy (in the form of tidal forces,
shear, rotation, etc.) also involves only the ratio of $\sigma^2$ to
$V_{\rm circ}^2$ (in the present approximation where background stars are
ignored). This ratio of binding energy scales inversely with the number of
star complexes.

It follows that morphologically clumpy galaxies produce thicker disks and
have more strongly bound star-forming regions than smooth galaxies.  The
thicker disks (for the same galaxy size) correspond to higher star
formation rates (see previous section), and the tighter self-binding
reasonably transforms to a higher fraction of star formation in the form of
bound clusters \citep{elmegreen11}. Greater self-binding should also
correspond to a more massive cluster at the upper end of the cluster mass
range, and possibly denser clusters of all mass. This shift toward a more
massive largest cluster corresponds to the formation of a super star
cluster at the upper end of a power law distribution of cluster masses. The
power law need not change as $\sigma/V_{\rm circ}$ increases, because that
is determined by turbulent and gravitational fragmentation, with some
contribution from sub-cluster coalescence, which are all scale-free
processes.

We can now see where the old globular clusters might have come from. The
metal-rich globular clusters, which tend to coincide with the disks and
bulges of today's galaxies, were presumably made in the giant clumps of
high-redshift clumpy galaxies \citep{kravtsov05,shap10}, which were also
massive enough to hold on to their winds and metals, and thereby achieve a
high metallicity. These clumpy galaxies are the predecessors of today's
giant spirals. The metal-poor globular clusters, which tend to coincide
with the halos of today's galaxies, presumably came from high redshift
dwarf galaxies \citep{elmegreen12}, which should also have been clumpy,
like today's dwarfs, but because of their inability to contain their
star-formation winds, lost the metal-enriched material that stellar
evolution generated. Here we are using the galaxy mass dependence of the
wind mass loading factor, $w$, discussed in section \ref{oversim}, to
explain the mass-metallicity relationship in galaxies
\citep{mannucci09,lilly13}.  Metal-poor globulars associated with disk
populations could have originated {\it in-situ} \citep{brodie14} when the
disk had a much lower mass and metallicity than today.

\section{Summary}

Lynden-Bell's wish of an ``oversimplified average law like Schmidt's
suggestion'' came true. The law is essentially SFR$\sim\epsilon\rho/t_{\rm
ff}$ or SFR$\sim M_{\rm gas}/\tau$ for $\tau\sim0.17t_{\rm H}$ with
instantaneous Hubble time $t_{\rm H}$. Because the gas depletion time
$\tau$ is relatively short compared to the accretion time, whole galaxies
can be in approximate equilibrium with the star formation rate equal to the
cosmic accretion rate and the metallicity approximately constant,
increasing only as the galaxy mass increases.

Locally, star formation is more dynamic than this, acting on a timescale of
several crossing times for each region and with an efficiency that is low
at the average density but increases with density, considering the
hierarchical distribution of mass and density during turbulent
fragmentation. This increasing efficiency explains how bound clusters can
form at high density when the average efficiency is low at low density. The
dependence of the efficiency on the shape of the density distribution
function may also explain how the clustering fraction in star formation
varies with environment, including an expected increase with $\sigma/V_{\rm
rot}$ from more tightly bound clouds.

Galaxies go through a sequence of equilibrium states while star-forming
regions inside galaxies change, move, and disrupt on dynamical times. The
difference is that galaxies have inert dark matter halos which hold the
star formation in place, while star-forming regions have nothing to hold
them down but disrupt by internal processes soon after they form. Both
systems accrete gas on dynamical timescales and return a high fraction of
this gas to the surrounding medium, but for galaxies, the returned gas
falls back into the dark matter potential, while for star formation, it
moves to the side and falls into another potential when the next cloud
forms.

\begin{figure}\epsscale{.9}
\plotone{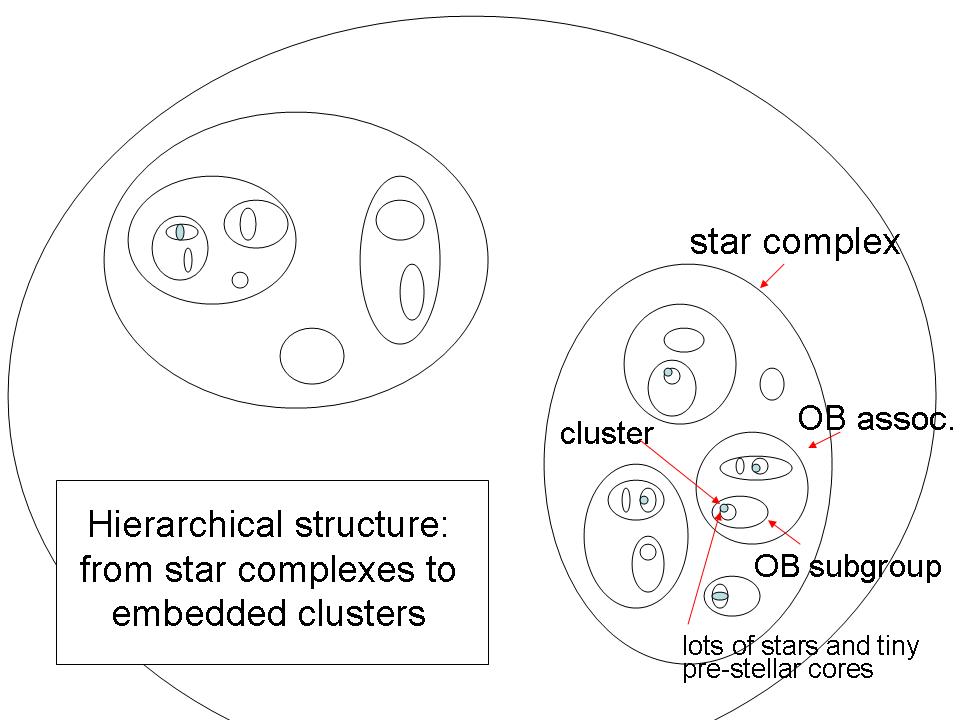}
\caption{Schematic showing hierarchical structure with smaller and younger star formation regions
nested inside larger and older star forming regions. As we zero in on the densest regions, the
low-density inter-region gas gets left behind and the average density increases. Also
increasing at the same time is the fraction of the gas at this high density that is at even higher
density. If the efficiency of star formation in some region is the mass fraction
of the densest sub-clumps (where stars form), then this mass fraction increases with
average density simply because of the hierarchical structure. The efficiency therefore increases
with average density.}
\label{fig1}\end{figure}

\end{document}